\documentclass[prd,twocolumn,showpacs,showkeys,floatfix]{revtex4}
\usepackage{amsmath,amssymb}
\usepackage{latexsym}
\usepackage{bm}

\usepackage{hyperref}
\usepackage[ansinew]{inputenc}
\makeatletter
\def\eqnarray{\stepcounter{equation}\let\@currentlabel=\theequation
\global\@eqnswtrue
\global\@eqcnt\z@\tabskip\@centering\let\\=\@eqncr
$$\halign to \displaywidth\bgroup\@eqnsel\hskip\@centering
  $\displaystyle\tabskip\z@{##}$&\global\@eqcnt\@ne
  \hfil${\;##\;}$\hfil
  &\global\@eqcnt\tw@ $\displaystyle\tabskip\z@{##}$\hfil
   \tabskip\@centering&\llap{##}\tabskip\z@\cr}
\makeatother

\begin{document}
\title{Non--static hyperbolically symmetric fluids}

\author{L. Herrera}
\email{lherrera@usal.es}
\affiliation{Departament de  F\'{\i}sica, Universitat Illes Balears, E-07122 Palma de Mallorca, Spain and Instituto Universitario de F\'isica
Fundamental y Matem\'aticas, Universidad de Salamanca, Salamanca 37007, Spain. \\
lherrera@usal.es}

\begin{abstract}
We present the general properties of  dynamic  dissipative fluid distribution endowed with hyperbolical symmetry. All the  equations required for its analysis are exhibited and used to contrast the behavior of the system  with the spherically symmetric case. Several exact solutions are exhibited and prospective applications to astrophysical and cosmological scenarios are discussed.
\end{abstract}
\date{today}

\maketitle

\section{INTRODUCTION}	
In a recent  paper  \cite{1} an alternative  global description of the  Schwarzschild black hole  has been proposed. The motivation behind such an endeavor  was, on the one hand    the fact that the space--time within the horizon, in the classical picture,  is necessarily non--static or, in other words, that  any transformation that maintains the static form of the Schwarzschild metric (in the whole space--time) is unable to remove the coordinate singularity appearing on the horizon in the line element  \cite{rosen}. 
Indeed, as  is well known,  no static observers can be defined inside the horizon (see \cite{Rin,Caroll} for a discussion on this point). This conclusion becomes intelligible if we recall  that the Schwarzschild horizon is also a Killing horizon, implying that the time--like Killing vector existing outside the horizon, becomes space--like inside it.

On the other hand, based on the physically reasonable point of view that  any equilibrium final state of a physical process should  be static, it would be desirable  to have  a static solution over the whole space--time.

Based on the arguments above, the following model was proposed in Ref. \cite{1}.

Outside the horizon ($R >2M$) one has the usual Schwarzschild  line element corresponding to the spherically symmetric vacuum solution to the Einstein  equations, which in polar coordinate reads (with signature $+2$)
\begin{eqnarray}
ds^2&=&-\left(1-\frac{2M}{R}\right)dt^2+\frac{dR^2}{\left(1-\frac{2M}{R}\right)}+R^2d\Omega^2, \nonumber \\ d\Omega^2&=&d\theta^2+\sin^2 \theta d\phi^2.
\label{w2}
\end{eqnarray}

This metric is static and spherically symmetric, meaning that it admits four Killing vectors:
\begin{eqnarray}
\mathbf{\xi}_{(\mathbf{0})} = \partial _{\mathbf{t}}, \quad {\bf \xi_{(2)}}=-\cos \phi \partial_{\theta}+\cot\theta \sin\phi \partial_{\phi},\nonumber \\
{\bf \xi_{(1)}}=\partial_{\phi}, \quad {\bf \xi_{(3)}}=\sin \phi \partial_{\theta}+\cot\theta \cos\phi \partial_{\phi}.
\label{2cmh}
\end{eqnarray}

The solution proposed for $R <2 M$  (with signature $(+ - - -)$) is
\begin{eqnarray}
ds^2&=&\left(\frac{2M}{R}-1\right)dt^2-\frac{dR^2}{\left(\frac{2M}{R}-1\right)}-R^2d\Omega^2, \nonumber \\ d\Omega^2&=&d\theta^2+\sinh^2 \theta d\phi^2.
\label{w3}
\end{eqnarray}

This is a static solution, meaning that it admits the time--like Killing vector  $\mathbf{\xi }_{(\mathbf{0})}$, however  unlike (\ref{w2})  it is not spherically symmetric, but hyperbolically symmetric, meaning that it admits  the three Killing vectors
\begin{eqnarray}
{\bf K_{(2)}}=-\cos \phi \partial_{\theta}+\coth\theta \sin\phi \partial_{\phi},\nonumber \\
{\bf K_{(1)}}=\partial_{\phi}, \quad {\bf K_{(3)}}=\sin \phi \partial_{\theta}+\coth\theta \cos\phi \partial_{\phi}.
\label{2cmhy}
\end{eqnarray}

Thus  the situation may be summarized as follows: if one wishes to keep sphericity within the horizon, one should abandon staticity, if one wishes to keep staticity within the horizon, one should abandon sphericity.

The classical picture of the black hole entails sphericity within the horizon, instead in \cite{1} we have proceeded differently and have assumed staticity within the horizon.

The three Killing vectors above define the hyperbolical symmetry.
Space--times endowed with hyperbolical symmetry have  previously  been the subject of research  in different contexts (see \cite{Ha}--\cite{pak3} and references therein).

In \cite{2nc} a general study of  geodesics in the spacetime described by (\ref{w3}) was presented, leading to some interesting conclusions about the behavior of  a test   particle in this new picture of the Schwarzschild  black hole, namely:
\begin{itemize}
\item the gravitational force inside the region $R<2M$ is repulsive.
\item test particles cannot reach the center.
\item test particles can cross the horizon outward, but only along the $\theta=0$ axis.
\end{itemize}

Motivated by these intriguing results we embarked into a general study of fluid distributions endowed with hyperbolical symmetry \cite{st1,hd,hd2}. 

It is the purpose of this paper to present the main results concerning dynamic and dissipative fluids endowed with hyperbolical symmetry. This   includes the general equations  governing the behavior of such fluids as well as  a selection of some exact analytical solutions.  We shall   discuss about  prospective applications of these results to the study of some astrophysical and cosmological problems. We shall not consider here the static case analyzed in \cite{st1}.

Finally, it is worth stressing the fact that although the main  motivation to undertake the study of fluids endowed with hyperbolical symmetry was the intriguing properties of  the black hole model briefly described above, the  results we are going to exhibit in the sections below are completely  independent on such a model.

\section{Basic equations and variables}
We consider  hyperbolically symmetric  distributions  of evolving
fluids, which may be  bounded, or not, from outside by a surface  $\Sigma^{e}$. On the other hand, as we shall see below, fluids endowed with hyperbolical symmetry cannot  fill the central region, therefore they should be  bounded from inside by a  surface   $\Sigma^{i}$.  The fluid is assumed to be locally anisotropic (principal stresses unequal) and undergoing dissipation in the form of heat flow (diffusion approximation).

Choosing comoving coordinates, the
interior metric, admitting the three Killing vectors (\ref{2cmhy}), may  be written as
\begin{equation}
ds^2=-A^2dt^2+B^2dr^2+R^2(d\theta^2+\sinh^2\theta d\phi^2),
\label{1}
\end{equation}
where  $A$, $B$ and $R$ are assumed
positive, and  due to the hyperbolical symmetry  are functions of $t$ and $r$. We number the coordinates $x^0=t$, $x^1=r$, $x^2=\theta$
and $x^3=\phi$.

The energy momentum tensor $T_{\alpha\beta}$ of the fluid distribution
may be written in its canonical form as
\begin{equation}
T_{\alpha \beta} = {\mu} V_\alpha V_\beta + P h_{\alpha \beta} + \Pi_{\alpha \beta} +
q \left(V_\alpha \chi_\beta + \chi_\alpha V_\beta\right) \label{Tab}
\end{equation}
with
$$ P=\frac{P_{r}+2P_{\bot}}{3}, \qquad h_{\alpha \beta}=g_{\alpha \beta}+V_\alpha V_\beta,$$

$$\Pi_{\alpha \beta}=\Pi\left(\chi_\alpha \chi_\beta - \frac{1}{3} h_{\alpha \beta}\right), \qquad
\Pi=P_{r}-P_{\bot}.$$

$$V^{\alpha}V_{\alpha}=-1, \;\; V^{\alpha}q_{\alpha}=0, \;\; \chi^{\alpha}\chi_{\alpha}=1,\;\;
\chi^{\alpha}V_{\alpha}=0$$.

$$
V^{\alpha}=A^{-1}\delta_0^{\alpha}, \;\;
q^{\alpha}=qB^{-1}\delta^{\alpha}_1, \;\;
\chi^{\alpha}=B^{-1}\delta^{\alpha}_1,
$$

where $\mu$,  $P_r$, 
$P_{\perp}$,  $q^{\alpha}$, $V^{\alpha}$ have the usual meaning, 
and $\chi^{\alpha}$ is unit four--vector along the radial direction.  It is worth noticing that  bulk or shear viscosity could be introduced by redefining 
 the radial and tangential pressures.  In addition, dissipation in the free streaming approximation can be absorbed in $\mu, P_r$ and $q$.

 Since the Lie derivative  $\mathcal{L}$ and the partial derivative commute, then
\begin{equation}
\mathcal{L}_K (R_{\alpha \beta}-\frac{1}{2}g_{\alpha \beta}{\cal R})=8\pi \mathcal{L}_KT_{\alpha \beta}=0,
\label{ccm1}
\end{equation}
implying that all physical variables only depend on $t$ and $r$.

Finally, it is worth mentioning that because of the hyperbolical symmetry there are only two possible unequal principal stresses  ($P_r, P_\bot$).
\subsection{Einstein equations and conservation laws}

The Einstein equations for (\ref{1}) and (\ref{Tab}) are
\begin{eqnarray}
  8\pi \mu&=& -\frac{1}{R^2}-\frac{1}{B^2}\left
  [-\frac{2B^\prime}{B}\frac{R^\prime}{R}+\left(\frac{R^\prime}{R}\right)^2+\frac{2R^{\prime\prime}}{R} \right ]\nonumber\\
   &+& \frac{1}{A^2}\left ( \frac{2\dot B}{B}\frac{\dot R}{R}+\frac{\dot R^2}{R^2} \right ),\label{EE00}
\end{eqnarray}

\begin{equation}\label{EE01}
 4\pi q=-\frac{1}{AB}\left ( \frac{R^\prime }{R}\frac{\dot{B}}{B}+\frac{A^\prime}{A}\frac{\dot
 R}{R}-\frac{\dot{R}^\prime}{R}\right),
\end{equation}

\begin{eqnarray}
8\pi P_r &=& \frac{1}{R^2}+\frac{1}{B^2}\left [\frac{2A^\prime}{A}\frac{R^\prime}{R}+\left(\frac{R^\prime}{R}\right)^2 \right
]\nonumber \\
   &+& \frac{1}{A^2}\left ( \frac{2\dot{A}}{A}\frac{\dot{R}}{R}-\frac{\dot R^2}{R^2}-\frac{2\ddot{R}}{R}\right
   ),\label{EE11}
\end{eqnarray}

\begin{eqnarray}
 8\pi P_\bot&=& \frac{1}{B^2}\left (
 -\frac{A^\prime}{A}\frac{B^\prime}{B}+\frac{A^\prime}{A}\frac{R^\prime}{R}-\frac{B^\prime}{B}\frac{R^\prime}{R}
 +\frac{A^{\prime\prime}}{A}+\frac{R^{\prime\prime}}{R} \right ) \nonumber\\
   &+& \frac{1}{A^2}\left (
   \frac{\dot{A}}{A}\frac{\dot{B}}{B}+\frac{\dot{A}}{A}\frac{\dot{R}}{R}-\frac{\dot{B}}{B}\frac{\dot{R}}{R}
   -\frac{\ddot B}{B}-\frac{\ddot R}{R}  \right ),\label{EE22}
\end{eqnarray}
where dots and primes denote derivative with respect to $t$ and $r$ respectively.
It is worth noticing the difference between these equations and the corresponding to the spherically symmetric case (see for example Eqs.(7)--(10) in \cite{epjc}).

The conservation laws $T^{\mu \nu}_{;\mu}=0$, as in the spherically symmetric case,  have only two independent components, which are displayed in the Appendix.
\subsection{Kinematical variables}

The four--acceleration $a_{\alpha}$ and the expansion $\Theta$ of the fluid are
given by
\begin{equation}
a_{\alpha}=V_{\alpha ;\beta}V^{\beta}, \;\;
\Theta={V^{\alpha}}_{;\alpha}. \label{4b}
\end{equation}
From  which we obtain  for the  four--acceleration  and its scalar $a$,
\begin{equation}
a_1=\frac{A^{\prime}}{A}, \;\; a=\frac{A^{\prime}}{AB}\Rightarrow a^\alpha=a\chi^\alpha, \label{5c}
\end{equation}
and for the expansion
\begin{equation}
\Theta=\frac{1}{A}\left(\frac{\dot{B}}{B}+2\frac{\dot{R}}{R}\right).
\label{5c1}
\end{equation}

The   shear tensor $\sigma_{\alpha\beta}$ is defined by (the vorticity vanishes identically)
\begin{equation}
\sigma_{\alpha\beta}=V_{(\alpha
;\beta)}+a_{(\alpha}V_{\beta)}-\frac{1}{3}\Theta h_{\alpha\beta},
\label{4a}
\end{equation}
its non zero components are
\begin{equation}
\sigma_{11}=\frac{2}{3}B^2\sigma, \;\;
\sigma_{22}=\frac{\sigma_{33}}{\sinh^2\theta}=-\frac{1}{3}R^2\sigma,
 \label{5a}
\end{equation}
and its scalar
\begin{equation}
\frac{3}{2}\sigma^{\alpha\beta}\sigma_{\alpha\beta}=\sigma^2,
\label{5b}
\end{equation}
reads
\begin{equation}
\sigma=\frac{1}{A}\left(\frac{\dot{B}}{B}-\frac{\dot{R}}{R}\right).\label{5b1}
\end{equation}

All the expressions above are the same, in terms of the metric functions, as in the spherically symmetric case.
\subsection{The Weyl tensor}
The magnetic part of the Weyl tensor for  our metric (\ref{1}) vanishes, whereas  its electric part may be written as

\begin{equation}\label{weyl1}
  E_{\alpha\beta}=\mathcal{E}\left(\chi_\alpha \chi_\beta-\frac{1}{3}h_{\alpha\beta}\right),
\end{equation}
with
\begin{eqnarray}
\mathcal{E}&=&\frac{1}{2B^2}\left[
-\frac{A^\prime}{A}\frac{R^\prime}{R}-\frac{A^\prime}{A}\frac{B^\prime}{B}+\frac{R^\prime}{R}\frac{B^\prime}{B}+
\left(\frac{R^\prime}{R}\right)^2+\frac{A^{\prime \prime}}{A}-\frac{R^{\prime \prime}}{R}\right]\nonumber \\
&+& \frac{1}{2A^2}\left[ \frac{\dot A}{A}\frac{\dot B}{B}-\frac{\dot A}{A}\frac{\dot R}{R}+\frac{\dot
R}{R}\frac{\dot B}{B}-
\left(\frac{\dot R}{R}\right)^2-\frac{\ddot B}{B}+\frac{\ddot R}{R}\right]+\frac{1}{2R^2}.\label{escE}
\end{eqnarray}

\subsection{The mass function}
Following \cite{Misner,Cahill}  we may define the mass function  as
\begin{equation}
m(r,t)=-\frac{R}{2}R^{3}_{232}=\frac{R}{2}\left[\left(\frac{R^{\prime}}{B}\right)^2-\left(\frac{\dot
R}{A}\right)^2+1\right],
 \label{fmasa}
\end{equation}
where the Riemann tensor component $R^{3}_{232}$ is now calculated for (\ref{1}).

Introducing the proper time derivative $D_T$, and the proper radial derivative $D_R$  by
\begin{equation}
D_T=\frac{1}{A}\frac{\partial}{\partial t}, \label{16}
\end{equation}
and
 \begin{equation}
D_R=\frac{1}{R^{\prime}}\frac{\partial}{\partial r},\label{23a}
\end{equation}
we may  define the ``areal'' velocity $U$ as
\begin{equation}
U=D_TR, \label{19}
\end{equation}
which must be smaller than $1$ (in relativistic units).

Then, since $U <1$, it follows at once from (\ref{fmasa}) that $m$ is a positive defined quantity.
Also, (\ref{fmasa}) can be rewritten as
\begin{equation}
E \equiv \frac{R^{\prime}}{B}=\left(\frac{2m}{R}+U^2-1\right)^{1/2}.
\label{20x}
\end{equation}

Using (\ref{fmasa})  with (\ref{16}), (\ref{23a}), and the field equations, we obtain

\begin{eqnarray}
D_Tm=4\pi\left(P_rU+ qE\right)R^2,
\label{22Dt}
\end{eqnarray}
and
\begin{eqnarray}
D_Rm=-4\pi\left(\mu+q\frac{U}{E}\right)R^2,
\label{27Dr}
\end{eqnarray}
producing
\begin{equation}
m=-4\pi\int^{r}_{0}\left( \mu +q\frac{U}{E}\right)R^2R^\prime dr, \label{27intcopy}
\end{equation}
satisfying the regular condition  $m(t,0)=0$.

Integrating (\ref{27intcopy}) we find
\begin{equation}
\frac{3m}{R^3} = -4\pi {\mu} +\frac{4\pi}{R^3} \int^r_0{R^3\left(D_R{ \mu}-3 q \frac{U}{RE}\right) R^\prime dr}.
\label{3m/R3}
\end{equation}

From (\ref{27intcopy})  two important general properties of hyperbolically symmetric fluids, follow.   

On the one hand, $\mu$ is necessarily negative. This follows from the (physically meaningful) condition that  the fluid is not very far from thermal equilibrium, implying $q<<\vert\mu\vert$, and   furthermore $R^\prime>0$ to avoid shell crossing, $m>0$ and $E$ is a regular function within the fluid.

 On the other hand, we see from  (\ref{27intcopy}) that whenever the energy density  is regular, then $m\sim r^3$ as $r$ tends to zero. However, in this same limit $U\sim 0$, and $R\sim r$ implying because of (\ref{20x}) that the central region  cannot be filled with our fluid distribution.  In other words, there is a central region of finite dimensions, which cannot be filled with a fluid endowed with hyperbolical symmetry. At this point many different scenarios may be envisaged. We shall assume here that the center is surrounded by a vacuum cavity described by a Minkowski space--time. It is worth emphasizing that such a choice does not affect the general properties of the fluids endowed with hyperbolic symmetry.

 The two above mentioned features of the fluid appear also in the static case \cite{st1}.

Before concluding this section it is worth  discussing with some detail  on equation (\ref{pprimaU2}), and compare it with the corresponding equation for the spherically symmetric case (see eq.(C6) in \cite{epjc}).

Equation (\ref{pprimaU2}) is particularly appealing because it  has  the ``Newtonian'' form
$Force = Mass \ density \times \ Acceleration$.  Let us now analyze its  different terms. The first term on the right represents the gravitational interaction,  it is the product of the passive gravitational mass density (p.g.m.d) ($\mu+P_r$), which due to the fact that the energy density is negative, would be negative,  and the active gravitational mass (a.g.m) ($4\pi P_r R^3-m$)  which would also be negative for most equations of state. Thus  the gravitational term has the same sign as in the spherically symmetric case. However its effect  is the inverse of this latter case. Indeed, since the p.g.m.d is negative so is the inertial mass density, implying that  the gravitational term tends to increase $D_TU$, i.e. it acts as a repulsive force, instead of  an attractive one as in (C6) of \cite{epjc}. Also,  we see that a negative pressure gradient which produces a force pointing outward, would tend to push any fluid element inwardly.

\section{THE TRANSPORT EQUATION}
Since we are considering dissipative systems we have to adopt a transport equation. In order to ensure causality we shall resort to   the  transport equation obtained form 
the  M\"{u}ller--Israel--Stewart  theory \cite{Is,Is2,Is3}.

Then, the  corresponding  transport equation for the heat flux reads

\begin{equation}
\tau
h^{\alpha\beta}V^{\gamma}q_{\beta;\gamma}+q^{\alpha}=-\kappa h^{\alpha\beta}
(T_{,\beta}+Ta_{\beta}) -\frac 12\kappa T^2\left( \frac{\tau
V^\beta }{\kappa T^2}\right) _{;\beta }q^\alpha ,  \label{21t}
\end{equation}
where $\kappa $  denotes the thermal conductivity, and  $T$ and
$\tau$ denote temperature and relaxation time respectively. 

There is only one non-vanishing independent component of  Equation (\ref{21t}),  which may be written as
\begin{equation}\label{ecTra}
  \tau D_Tq=-q-\frac{\kappa}{AB}(AT)^\prime-\frac{1}{2}\tau \Theta q-\frac{1}{2}\kappa T^2D_T\left(\frac{\tau}{\kappa T^2}\right)q.
\end{equation}
In the case $\tau=0$ we recover the Eckart--Landau equation \cite{17T}.

Under some circumstances it is possible to adopt  the so called ``truncated'' version where the last term in (\ref{21t}) is neglected \cite{PAN},
\begin{equation}
\tau
h^{\alpha\beta}V^{\gamma}q_{\beta;\gamma}+q^{\alpha}=-\kappa h^{\alpha\beta}
(T_{,\beta}+Ta_{\beta}) \label{V1},
\end{equation}
and whose    only non--vanishing independent component becomes
\begin{equation}
\tau \dot q+qA=-\frac{\kappa}{B}(TA)^{\prime}. \label{V2}
\end{equation}

Two important thermodynamical properties of  hyperbolically symmetric fluids can be inferred from (\ref{ecTra}).

Let us first consider the condition of thermal equilibrium.

As it was pointed out by Tolman many years ago \cite{Tol}, the condition of thermal equilibrium in the
presence of a gravitational field must change with respect to its form in the absence
of gravity since  thermal energy tends to displace to regions of lower gravitational
potential  (independently on any temperature gradient).  Thus, a temperature gradient is necessary in thermal equilibrium in order
to prevent the flow of heat from regions of higher to lower gravitational potential.

Indeed, as it follows at once from (\ref{ecTra}), thermal equilibrium implies 
\begin{equation}
\left(TA\right)^\prime=0 \Rightarrow T^\prime=-\frac{T}{A}A^\prime=-TaB.
\label{t1}
\end{equation}
However in our case $a<0$, (the four--acceleration  is now directed radially inwardly), implying the existence of a repulsive   gravitational force, leading to a positive temperature gradient in order to assure thermal equilibrium. This situation is at variance with the spherically symmetric case, where a negative temperature gradient is required to assure thermal equilibrium.

A second interesting property appears from the combination of (\ref{pprimaU2}) with (\ref{ecTra}), producing (\ref{pprimaU3}).  It brings out the effect of dissipative processes  on the p.g.m.d., and by virtue of the equivalence principle, on the effective inertial mass density as well. This kind of effect was pointed out for the first time for the spherically symmetric case in \cite{1tef} (see also \cite{ther}  for a discussion on this effect). In our case the term $\frac{\kappa T}{\tau}$ increases the absolute value of the effective p.g.m.d (which is negative), thereby increasing the absolute value of the effective inertial mass density (the term in the bracket on the left of (\ref{pprimaU3})), as a result of which  any hydrodynamic force directed outward tends to push  the fluid element  inward but  weaker  than in the non--dissipative case, due to the term $\frac{\kappa T}{\tau}$. Thus in the hyperbolically symmetric  case the thermal effect on the inertial mass density enhances the tendency to expansion,  as in the spherically symmetric case, but comes about  from different way.

In order to obtain specific solutions to the Einstein equations we shall need to impose additional restrictions. In this work we shall assume two different types of restrictions.  On the one hand, we shall assume that the fluid evolves in the quasi--homologous regime and satisfies the vanishing complexity factor condition. The next  section is devoted to explain these conditions in some detail. On the other hand we shall seek for spacetimes which could be considered as hyperbolically symmetric versions of Lemaitre--Tolman--Bondi spacetimes, for doing so we shall assume that the fluid is geodesic, shearing, non--conformally flat and the energy density is inhomogeneous.

\section{COMPLEXITY FACTOR AND QUASI--HOMOLOGOUS EVOLUTION}
The complexity factor is a scalar function intended to measure the degree of complexity of a self-gravitating system  (in some cases more than one scalar function may be required).
For a static, hyperbolically symmetric fluid distribution it was assumed in \cite{st1} (following the arguments developed in \cite{c1}) that the simplest system corresponds  to a 
 homogeneous (in the energy density), locally  isotropic fluid distribution (principal stresses equal). Thus, a zero value of  the
 complexity factor is assigned  to  such a distribution. Furthermore, it was shown that   a single scalar function (hereafter referred to as $Y_{TF}$) describes the modifications introduced by the energy density inhomogeneity and pressure anisotropy   to the Tolman mass, with respect to its value  for the vanishing  complexity factor.

This  scalar belongs to   a set of variables named  structure scalars, defined in \cite{20} and which appear in the orthogonal splitting of the Riemann tensor \cite{18,19,20}.  For our purpose here we shall need only one of the five structure scalars characterizing   our fluid distribution, namely  $Y_{TF}$ .

The scalar $Y_{TF}$  defines the trace--free part of the  electric Riemann tensor  $Y_{\alpha\beta}$  given by 

\begin{eqnarray}
Y_{\alpha\beta} = R_{\alpha \gamma \beta \delta}V^\gamma
V^\delta.\label{Telectric} \end{eqnarray}

In our case the expression for $Y_{TF}$ reads (see \cite{20} for details)

\begin{eqnarray}
Y_{TF}={\cal E}-4\pi \Pi.\label{TEY}
\end{eqnarray}

Using (\ref{EE11}), (\ref{EE22}) and  (\ref{escE}), we may express $Y_{TF}$ in terms of the metric functions and their derivatives as

\begin{eqnarray}
  Y_{TF}&=&\frac{1}{B^2}\left (\frac{A^{\prime\prime}}{A}-\frac{A^\prime}{A}\frac{R^\prime}{R}-\frac{A^\prime}{A}\frac{B^\prime}{B}\right)\nonumber\\
 &+& \frac{1}{A^2}\left (\frac{\dot A}{A}\frac{\dot B}{B}- \frac{\dot A}{A}\frac{\dot R}{R} -\frac{\ddot{B}}{B}+\frac{\ddot{R}}{R}\right ).\label{ecYTF}
\end{eqnarray}

In the dynamic case, we still need to provide a criterion for the definition of complexity  of the pattern of evolution.

 We  shall consider the  quasi-homologous evolution defined in \cite{epjc} as the simplest mode of evolution.

The quasi--homologous condition  reads
\begin{equation}
U=R\frac{U_{\Sigma^e}}{R_{\Sigma^e}},
 \label{qsh4bis}
 \end{equation}
implying

 \begin{equation}
\frac{4\pi q }{E}+\frac{\sigma}{R}=0.
 \label{qsh4}
 \end{equation}
 
The above condition will be used to obtain specific models, the rationale behind such a choice is based on the fact that it represents  the relativistic version of the well-known homologous condition widely used in classical  astrophysics, furthermore as shown in \cite{epjc},  it qualifies as   one of the simplest  patterns of evolution.

\section{JUNCTION CONDITIONS}
In the case that the fluid is bounded then junction conditions on the boundary have to be imposed \cite{Darmois}, otherwise   we have to cope with  the presence of thin shells \cite{17}.

Thus, outside $\Sigma^e$  we assume that we have the hyperbolic version of the Vaidya
spacetime, described by
\begin{equation}
ds^2=-\left[\frac{2M(v)}{\bf r}-1\right]dv^2-2 d{\bf r} dv+{\bf r}^2(d\theta^2
+\sinh^2\theta
d\phi^2) \label{1int},
\end{equation}
where $M(v)$  denotes the total mass,
and  $v$ is the retarded time.

The continuity of the first and the second fundamental forms imply, on 
$\Sigma^e$, 
\begin{equation}
q\stackrel{\Sigma^e}{=}P_r.\label{j3}
\end{equation}
 and 
 \begin{equation}
m(t,r)\stackrel{\Sigma^e}{=}M(v), \label{junction1}
\end{equation}
where $\stackrel{\Sigma^e}{=}$ means that both sides of the equation
are evaluated on $\Sigma^e$.

In the cases where the central region is surrounded by an empty vacuole bounded by a surface $\Sigma^i$, junction conditions at the inner boundary of the fluid distribution read
\begin{equation}
P_r\stackrel{\Sigma^i}{=}0,\label{j3b}
\end{equation}
and
 \begin{equation}
m(t,r)\stackrel{\Sigma^i}{=}0. \label{junction1b}
\end{equation}
\section{SOME EXACT SOLUTIONS}

In the following subsections we shall exhibit several families of solutions to the Einstein equations for hyperbolically symmetric fluids. More specifically, we shall consider  two families of solutions. One of them  will be obtained  by assuming quasi--homologous evolution and the vanishing of the complexity factor. The other family of solutions corresponds to hyperbolically symmetric versions of Lemaitre--Tolman--Bondi (LTB) space--times.  In each case dissipative and non--dissipative models were obtained. The purpose of the presentation of these models is not only the  potential application of some of them to the study of specific astrophysical scenarios, but also to  illustrate the richness of fluid distributions endowed with hyperbolical symmetry.

\subsection{ Models with vanishing complexity factor and evolving in the quasi--homologous regime}

In this case we shall exhibit only non--dissipative models (for dissipative models see \cite{hd}). Then, excluding dissipative processes, and assuming the quasi--homologous condition (\ref{qsh4}) we may write

\begin{equation}\label{qcero}
  q=0\,\,\Rightarrow\,\, \sigma=0\,\,\Rightarrow\,\, \frac{\dot B}{B}=\frac{\dot R}{R}\,\,\Rightarrow\,\, R=rB,
\end{equation}
and

\begin{equation}\label{chomo}
  U=\frac{\dot R}{A}=\frac{r\dot B}{A}={\tilde a}(t)rB.
\end{equation}
Imposing next the condition  $Y_{TF}=0$ we have

\begin{equation}\label{YTFcero1}
  \frac{A^{\prime\prime}}{A}-\frac{A^\prime}{A}\frac{B^\prime}{B}-\frac{A^\prime}{A}\frac{R^\prime}{R}=0.
\end{equation}

In order to exhibit specific  solutions, we shall further assume some additional restrictions.

\subsubsection{ $\mathcal{E}=0$, $\Pi=0$ }
We shall assume here that the fluid is conformally flat ($\mathcal{E}=0$) and the pressure is isotropic ($\Pi=0$), which combined with $Y_{TF}=0$ produces $\mu^\prime=0$ (i.e. the energy density is homogeneous).

In this case the metric functions are (see \cite{hd} for details)
\begin{eqnarray}\label{nds1}
  R &=& \frac{\tilde R(t)}{\cos[c_1(t)+\ln r]}, \\
  B &=&  \frac{\tilde R(t)}{r \cos[c_1(t)+\ln r]},  \\
  A &=&\gamma(t) \tilde R^2(t) \tan [c_1(t)+\ln r]+b(t), \label{nds7}
\end{eqnarray}
where $\tilde R(t), c_1(t), \gamma(t), b(t)$ are arbitrary functions of their argument. 

We shall further  specify the  solution by choosing the above functions as follows

\begin{equation}\label{condFRW}
  \dot c_1=\frac{\dot{\tilde R}}{\tilde R}, \qquad b(t)=\gamma(t)\tilde R^2,
\end{equation}
producing

\begin{eqnarray}
  \frac{\dot R}{R} &=& \frac{\dot{\tilde R}}{\tilde R}(1+\tan u), \\
  A &=& \gamma(t)\tilde R^2(1+\tan u),\quad \Rightarrow \quad A= \frac{\tilde a\dot R}{R}\label{condA},
\end{eqnarray}
with  $\tilde a=\frac{\gamma(t)\tilde R^3}{\dot{\tilde R}}$ and  $u=c_1(t)+\ln r$.
From  the above expressions we found for the physical variables and the mass function,

\begin{eqnarray}
  8\pi \mu &=& -\frac{3}{\tilde R^2}+\frac{3}{\tilde a^2}, \\
  8\pi P_r =8\pi P_\bot&=&-\frac{3}{\tilde a^2}+\frac{3\tan u+1}{\tilde R^2(\tan u+1)}\nonumber 
  \\ &+&\frac{2\tilde R \dot{\tilde a}}{\tilde a^3 \dot{\tilde R}(\tan u+1)},\label{nds2}
  \\
  m &=& \frac{\tilde{R}}{2 \cos ^3 u}\left(1-\frac{\tilde {R}^2}{\tilde{a}^2}\right).
\end{eqnarray}

If we choose  $\tilde R(t), c_1(t), \gamma(t)$ such that they tend to a constant as $t\rightarrow \infty$, then the above solution tend to the incompressible isotropic solution found in  \cite{mimc2}, which is a particular case of the hyperbolically symmetric Bowers--Liang solution found in \cite{st1}.

This model  might be considered as a hyperbolically symmetric version of the Friedman--Robertson--Walker space--time (FRW), since both  share some  similar properties e.g. $\mathcal{E}=\Pi=\mu^\prime=\sigma=0$. However our solution  is not geodesic as in the spherically symmetric case. We shall next find another  version of the hyperbolically symmetric  FRW space--time, but satisfying the geodesic condition $A^\prime=0$.
\subsubsection{$A=1$, $\mathcal{E}=0$}

If we further impose the geodesic condition on the fluid ($A=1$),  then  the quasi--homologous condition implies
\begin{equation}
\frac{R_I}{R_{II}} = constant,
\label{g1}
\end{equation}
where $R_I$ and $R_{II}$ denote the areal radii of two shells $(I, I I)$ described by $r = r_I = constant$, and $r = r_{II} = constant$, respectively. In the notation of \cite{hd},  conditions (\ref{chomo}) and (\ref{g1}) define the homologous evolution.

From (\ref{g1}) it follows at once that  $R$ is a separable function.

The conditions $A=1$ and  $q=0$  imply

\begin{equation}\label{ecA1q0}
\frac{\dot{B}}{B}=\frac{\dot{R}^\prime}{R^\prime},
\end{equation}
where (\ref{EE01}) has been used.
From the shear--free  condition  and the separability of $R$ it can be easily shown that $B$ becomes a function of $t$ alone $B=B(t)$, i.e.
\begin{equation}\label{YTF0A1}
 R=rB(t).
\end{equation}
Then (\ref{ecA1q0}) is automatically satisfied,  as well as  $Y_{TF}=0$ as it follows from  (\ref{YTFcero1}).

The physical variables and the mass function  for this model read
\begin{eqnarray}
8\pi\mu&=&-\frac{2}{r^2 B^2}+\frac{3 \dot{B}^2}{B^2},
\label{1b}\\
8\pi P_r&=&\frac{2}{r^2 B^2}-\frac{\dot{B}^2}{B^2}-\frac{2\ddot{B}}{B},
\label{2b}\\
8\pi P_{\bot}&=&-\frac{2\dot{B}^2}{B^2}-\frac{2\ddot{B}}{B},
\label{3b}\\
m&=&\frac{rB}{2}(2-r^2 \dot {B}^2).
\label{4b}
\end{eqnarray}
Thus the fluid  is conformally flat, shear--free, geodesic, evolves homologously and satisfies the  vanishing complexity factor condition. So, it also qualifies as a hyperbolically symmetric version of FRW space--time.  However, unlike the spherically symmetric case, it is anisotropic in the pressure and the energy--density is inhomogeneous.

It is worth analyzing with some detail the differences between this case and the situation in  the spherically  symmetric case (FRW). In the latter case we have seen \cite{c2} that for a non--dissipative fluid satisfying the homologous condition, the complexity factor vanishes and there is a single solution characterized by $\Pi=\mu^\prime=a={\cal E}=0$ (FRW).

However in the present case, imposing homologous condition on a geodesic non--dissipative fluid we get a conformally flat, shear--free fluid  with  $\Pi, \mu^\prime\neq 0$.
If we want to describe an isotropic, homogeneous, shear--free non--dissipative  fluid, then we have to relax the geodesic condition.

Finally,  let us build a toy model with the above solution, by choosing a particular form of  $B$ such that asymptotically it leads to a static regime. 

Thus, let us assume.
\begin{equation}
B=\beta\left(1+e^{-\alpha t}\right),
\label{t1}
\end{equation}
where $\alpha, \beta$ are two positive constants.

Then it  is a simple matter to check that as $t\rightarrow \infty$ we get
\begin{eqnarray}
  8\pi \mu &=& -\frac{2}{r^2 \beta^2},
  \\
  8\pi P_r &=& \frac{2}{r^2 \beta^2},\\
  8\pi P_\bot &=&0,
\end{eqnarray}
and for the mass function we get asymptotically $m=r\beta$.

As we see, our toy model converges to the static solution corresponding to the stiff equation of state $(P_r=\vert \mu \vert)$ found in \cite{st1} (Eqs.(138-139) in that reference).

\subsection{Hyperbolically symmetric versions of Lemaitre--Tolman--Bondi spacetimes}
We shall now consider a family of solutions sharing some basic properties with LTB space--times, namely.

\begin{itemize}

\item $\mu^\prime, \sigma, {\cal E} \neq 0$, $A=1$.
\item We shall NOT require homologous condition.

\end{itemize}
Let us first consider the simplest case (non--dissipative dust) and afterward we shall consider extensions to the dissipative anisotropic case.
\subsubsection{The non--dissipative dust case}

In this case it follows from the field equations
\begin{equation}
B(t,r)=\frac{R^\prime}{\left[k(r)-1\right]^{1/2}},\label{BTBh}
\end{equation}
\noindent where $k$ is an arbitrary function of $r$.

Next, from the definition of the mass function we may write
\begin{equation}\dot R^2=-\frac{2m}{R}+k(r),
\label{intltb1h}
\end{equation}

 implying  $k(r)>\frac{2m}{R}$. Thus unlike the spherically symmetric LTB space--time we now have only one case $k(r)>0$.

The solution to (\ref{intltb1h}) may be written as:

\begin{equation}
 R=\frac{m}{k}(\cosh \eta+1), \qquad  \frac{m}{k^{3/2}}(\sinh \eta+\eta)=t-t_{0}(r),
\label{int5h}
\end{equation}
and the line element reads
\begin{equation}
ds^2=-dt^2+\frac{(R^{\prime})^2}{k(r)-1}dr^2+R^2(d\theta^2+\sinh^2\theta
d\phi^2).\label{mTB}
\end{equation}

In order to prescribe an explicit model we have to provide the three functions $k(r)$, $m(r)$ and $t_{0}(r)$. However, since (\ref{mTB}) is invariant under  transformations of the form $r=r(\tilde r)$, we only need two functions of $r$.

Assuming  as example $m_0=\frac{m}{k}=constant$, and $t_{0}(r)=constant$ the expressions for $\Theta$ and  $\sigma$ read
\begin{eqnarray}
  \Theta &=& \frac{\sqrt{k}}{m_0}\left ( \frac{\sinh \eta}{\sinh \eta +\eta}+\frac{\cosh \eta+2\sinh\eta+1}{(\cosh \eta+1)^2} \right ), \\
  \sigma &=& \frac{\sqrt{k}}{m_0}\left ( \frac{\sinh \eta}{\sinh \eta +\eta}+\frac{\cosh \eta-\sinh\eta+1}{(\cosh \eta+1)^2} \right ),
\end{eqnarray}
from where it is clear that the expansion is always positive.

The only non--trivial conservation law in this case reads
\begin{equation}
\dot \mu+\mu \Theta=0,
\label{cl1}
\end{equation}
or

\begin{eqnarray}
\dot \mu+\mu\left(\frac{\dot B}{B}+2\frac{\dot R}{R}\right)=0,\label{an'}
\end{eqnarray}
producing
\begin{equation}
 \mu=\frac{h(r)}{BR^2},\label{nltb}
\end{equation}
or, using (\ref{BTBh})
\begin{equation}
 \mu=\frac{3 h(r)\left[k(r)-1\right]^{1/2}}{(R^{3})^\prime},\label{n1ltb}
\end{equation}
where $h(r)$ is a function of integration, which due to the fact that the energy density is negative, must be necessarily negative. For this family of solutions it can  be shown the the complexity factor cannot vanish.

\subsubsection{Anisotropic dissipative models}

We shall now consider the possibility that the system radiates, and the pressure is non--vanishing and may be anisotropic.  Then following the scheme developed in  \cite{LTBd}, we write
\begin{equation}
B(t,r)=\frac{R^\prime}{\sqrt{K(t,r)-1}},\label{ecB}
\end{equation}
where
\begin{equation}\label{K1}
  K(t,r)-1=\left [ \int 4\pi {q}Rdt  +C(r) \right ]^2,
\end{equation}
and $C(r)=\sqrt{k(r)-1}$.

Thus, the general form of the line element may be written as 
\begin{equation}
  ds^2=-dt^2+\frac{(R^\prime)^2 dr^2}{\left [ \int 4\pi {q}Rdt  +C(r) \right ]^2}+R^2(d\theta^2 +\sinh ^2\theta d\phi^2).
\end{equation}

A first class of models of this family will be obtained from conditions on the complexity factor, while a second one  will be obtained by a specific restriction on the function $B$, particularly suitable for describing situations  where a cavity surrounding the central region appears \cite{16n}.

Let us first consider models obtained upon conditions on the complexity factor.

In our case the complexity factor $Y_{TF}$ may be written as 
\begin{eqnarray}
  Y_{TF} &=& \frac{\ddot{R}}{R}-\frac{\ddot{R}^\prime}{R^\prime} +\frac{\ddot{K}}{2(K-1)}\nonumber \\
  &+&\frac{\dot{K}}{K-1}\left (\frac{\dot{R}^\prime}{R^\prime}-\frac{3}{4}\frac{\dot K}{K-1} \right)
\end{eqnarray}
\noindent In order to obtain specific models we shall assume that the complexity factor  has the same form as in the non-dissipative case, implying
\begin{equation}
\frac{\ddot{K}}{2(K-1)}
  +\frac{\dot{K}}{K-1}\left (\frac{\dot{R}^\prime}{R^\prime}-\frac{3}{4}\frac{\dot K}{K-1} \right)=0.\label{ecK}
\end{equation}

\noindent The integration of  (\ref{ecK}) produces

\begin{equation}
\frac{R^\prime \sqrt{\dot{K}}}{(K-1)^{\frac{3}{4}}}=C_1(r)\label{pik},
\end{equation}
where $C_1$ is an integration function.

From the above, using the field equations we may write for $q$
\begin{equation}\label{ecK4}
  2\pi {q}=\frac{1}{R\left ( R^\prime C_1(r) \int \frac{dt}{(R^\prime)^2}\right)^2}.
\end{equation}

Now, it is worth noticing that this class of solutions does not admit pure dust condition (pressure cannot vanish).

Indeed, from the dust condition and (\ref{ecK}) we obtain

\begin{equation}\label{K1122}
 \frac{\dot{R}}{R} \frac{C_1^2\sqrt{K-1}}{2(R^\prime)^2}=0,
\end{equation}

\noindent which cannot be satisfied, implying that there are no radiating dust solutions in this case. Therefore we have to relax the dust condition, and we shall consider models of radiating anisotropic fluids. A simple  model of this kind, may be obtained by assuming
{$P_\bot=0$,  $P_r\neq 0$ }.

 Then, from the condition $P_\bot=0$, we obtain 

\begin{equation}\label{BPbot}
  \ddot{B}=0,\quad \Rightarrow B=b_1(r) t+b_2(r),
\end{equation}
  where $b_1$ and $b_2$ are two arbitrary functions.

Using  (\ref{ecB}) we find for  $R$
\begin{equation}\label{RPbot}
  R^\prime-B\dot{R}=0, \quad \Rightarrow\quad R=\Phi\left[a_1(r) t+a_2(r)\right],
\end{equation}
where $\Phi$ is an arbitary function of its argument
\noindent and
\begin{equation}
a_1(r)=e^{\int b_1(r)dr},\qquad a_2(r)=\int b_2(r)e^{\int b_1(r)dr}dr.\label{a1a2}
\end{equation}

For specifying further the model let us choose  $b_1(r)$ and $b_2(r)$ as

\begin{equation}\label{bb1}
  b_1(r)=\frac{\beta_1}{r+\beta_2},\qquad b_2(r)=(r+\beta_2)^\alpha,
\end{equation}

\begin{eqnarray}
  B &=& \frac{\beta_1 t}{r+\beta_2}+(r+\beta_2)^{\alpha},\\
  R &=& (a_1t+a_2)^n,
\end{eqnarray}
where $\beta_1$, $\beta_2$, $\alpha$ and $n$ are arbitrary constants, chosen positive to ensure the positivity of $B$ and $R$.

Then  the physical variables and the complexity factor for this model read

\begin{eqnarray}
  8\pi \mu &=&-\frac{1}{(a_1t+a_2)^{2n}} -\frac{2(n-1)na_1^2}{(a_1t+a_2)^2},\label{d2}\\
  4\pi q &=& \frac{(n-1)na_1^2}{(a_1t+a_2)^2},\label{qn1} \\
  8\pi P_r &=&\frac{1}{(a_1t+a_2)^{2n}} -\frac{2(n-1)na_1^2}{(a_1t+a_2)^2},\label{p2}\\
m&=&\frac{(a_1t+a_2)^n}{2},
\end{eqnarray}

\begin{eqnarray}
  \Theta &=& \frac{b_1}{b_1t+b_2}+\frac{2n a_1}{a_1t+a_2} ,\\
 \sigma&=& \frac{b_1}{b_1t+b_2}-\frac{n a_1}{a_1t+a_2},
\end{eqnarray}

\begin{equation}
Y_{TF}=\frac{(n-1)na_1^2}{(a_1t+a_2)^2}.
\label{cf1}
\end{equation}
It is worth noticing that $Y_{TF}$ has exactly the same expression as $q$, as given by (\ref{qn1}). Therefore any solution of this family satisfying the vanishing complexity factor is necessarily non--dissipative.
On the other hand  $Y_{TF}$  is zero  if  $n=1$. Thus the solution of this family with vanishing complexity factor is characterized by $n=1$, which using (\ref{d2}) and (\ref{p2}) produces
\begin{equation}
P_r=-\mu.
\label{st}
\end{equation}
Finally, let us mention that the the expansion scalar is always positive.

We shall next obtain some models by imposing  a specific condition on $B$ ($B=1$). The reason to assume such a condition comes from the fact that, as shown in \cite{16n}, it is particularly suitable for describing situations where a vacuum cavity surrounds the central region. In this case the field equations read

\begin{eqnarray}\label{EEB1}
 8\pi  \mu &=& -\frac{1}{R^2}-\frac{2 R^{\prime\prime}}{R}-\left(\frac{R^\prime}{R}\right)^2+\frac{\dot{R}^2}{R^2}, \\
  4\pi q &=& \frac{\dot{R}^\prime}{R}, \label{101}\\
  8\pi  P_r &=& \frac{1}{R^2} +\left(\frac{R^\prime}{R}\right)^2-\left [\left(\frac{\dot R}{R}\right)^2+\frac{2\ddot{R}}{R}\right],\label{102}\\
 \end{eqnarray}
\begin{eqnarray}  
  8\pi P_\bot &=&  \frac{R^{\prime\prime}}{R}-\frac{\ddot{R}}{R}.\label{103}
\end{eqnarray}

Let us first consider  the non--dissipative case ($q=0$). Then it follows at once from (\ref{101}) that 

\begin{equation}
R=R_1(t)+R_2(r),
\label{se1}
\end{equation}
where $R_1$ and $R_2$ are arbitrary functions of their  arguments.

In order to exhibit an exact solution let us further assume $P_\bot=0$. Using this condition in (\ref{103}) produces
\begin{equation}
R_1(t)=a_1 t^2+b_1t +c_1,\qquad R_2(r)=a_1 r^2+b_2 r+c_2,
\label{se3}
\end{equation}
where $a_1, b_1, c_1, b_2, c_2$ are positive arbitrary constants.

The physical and kinematical variables for this model are
\begin{equation}
8\pi \mu=\frac{1}{\alpha^2}\left(-1-4a_1 \alpha-\beta^2+\gamma^2\right),
\label{se4}
\end{equation}

\begin{equation}
8\pi P_r=\frac{1}{\alpha^2}\left(1-4a_1 \alpha+\beta^2-\gamma^2\right),
\label{se5}
\end{equation}

\begin{equation}
P_r+\mu=-\frac{a_1}{\pi \alpha},
\label{semp}
\end{equation}

\begin{equation}
\Theta=\frac{2 \gamma}{\alpha}\label{se6},
\end{equation}

\begin{equation}
\sigma=-\frac{\gamma}{\alpha}\label{se7},
\end{equation}

\begin{equation}
m=\frac{ \alpha}{2}\left(\beta^2-\gamma^2+1\right),\label{se8}
\end{equation}

where
\begin{eqnarray}
\alpha&\equiv& a_1(t^2+r^2)+b_1t+b_2r+c_1+c_2;\qquad \beta\equiv 2a_1r+b_2, \nonumber \\ &&\gamma\equiv 2a_1t+b_1.
\label{se9}
\end{eqnarray}

For this model the expression for $Y_{TF}$ reads
\begin{equation}
Y_{TF}=\frac{2a_1}{\alpha}.
\label{cf3}
\end{equation}

Therefore the vanishing complexity factor condition implies $a_1=0$, producing because  of (\ref{se4})  and (\ref{se5})
\begin{equation}
P_r=-\mu.
\label{st3}
\end{equation}
Thus the solution of this family with the vanishing complexity factor  condition is also characterized by the stiff equation of state. Besides, the expansion scalar is always positive.

Finally, let us now consider the dissipative case ($q\neq0$). If we impose the condition $P_\bot=0$, then we get the equation $\ddot R=R^{\prime \prime}$, whose general solution is of the form
\begin{equation}
R(t,r)=c_1 \Psi(t+r) +c_2\Phi(t-r),
\label{se10}
\end{equation}
where $c_1, c_2$ are arbitrary constants, and $\Psi, \Phi$ arbitrary functions. 

As an example let us choose

\begin{equation}
R=a_1(t-r)^n,
\label{se18}
\end{equation}
where $a_1, n$ are positive constants, and the solution applies in the region $t>r$.

The ensuing physical and kinematical variables are in this case:
\begin{equation}
8\pi \mu=-\frac{1}{a_1^2\left(t-r\right)^{2n}}-\frac{2n\left(n-1\right)}{\left(t-r\right)^2},
\label{se19}
\end{equation}
\begin{equation}
4\pi q=-\frac{n\left(n-1\right)}{\left(t-r\right)^2},
\label{se20}
\end{equation}

\begin{equation}
8\pi P_r=\frac{1}{a_1^2\left(t-r\right)^{2n}}-\frac{2n\left(n-1\right)}{\left(t-r\right)^2},
\label{se21}
\end{equation}
\begin{equation}
\Theta=\frac{2 n}{t-r}, \quad \sigma=-\frac{ n}{t-r},
\label{se22}
\end{equation}

\begin{equation}
m=\frac{a_1\left(t-r\right)^n}2,
\label{se24}
\end{equation}

\begin{equation}\label{T2}
  T(t,r)=\frac{n(n-1)}{4\pi \kappa (t-r)}-\frac{n(n-1)\tau}{4\pi \kappa (t-r)^2}+T_0(t),
\end{equation}

where the temperature has been calculated used the truncated transport equation (\ref{V2}).

The corresponding expression of the complexity factor for this case reads
\begin{equation}
Y_{TF}=\frac{n(n-1)}{(t-r)^2}.
\label{cf5}
\end{equation}

Thus, the vanishing complexity factor  conditions requires $n=1$, implying because of (\ref{se20}) that the fluid is non--dissipative, and because of (\ref{se19}) and (\ref{se21}) that the fluid satisfies the stiff equation of state $P_r=-\mu$. 

As in the previous model, the expansion scalar is always positive.

\section{CONCLUSIONS}
The  general approach hereby described to analyze  the dynamics of hyperbolically symmetric fluids, including dissipative processes, brings out some  remarkable features of hyperbolically symmetric fluids, namely:
\begin{enumerate}
\item The energy density is necessarily negative.
\item The Tolman condition for thermodynamic equilibrium implies in this case the presence of a positive temperature gradient, unlike the negative temperature gradient required in the spherically symmetric case. 
\item The  thermal modification of  the inertial mass density  reported for the spherically symmetric case in \cite{1tef}, produces an effect that is similar   to the one  obtained in the spherically symmetric case (to  enhance the tendency to expansion) but  comes about through  the increasing of the absolute value of the effective inertial mass density.
\item The fluid cannot fill the central region.
\end{enumerate}

The violation of the weak energy condition ($\mu<0$) should not scare us. Indeed,  while it is true that at classical level we do not expect negative energy density in a realistic fluid, the situation is quite different in presence of  quantum effects. In fact there is abundant theoretical evidence illustrating the appearance of negative energy density  in some astrophysical and cosmological  scenarios  (see \cite{we1,we2,we3,cqg,pav} and references therein). Thus the type of fluids considered in this manuscript  might be useful for studying systems under extreme conditions where quantum effects are expected to play a relevant role.

This negative energy density implies the appearance of a repulsive  gravitational force which has two important thermodynamic consequences mentioned in points 2 and 3 above.

Finally, the fact that  the fluid distribution cannot  fill the central region is consistent with the result obtained in \cite{2nc}, indicating that  test particles are not allowed to reach the center for the line element (\ref{w3}). Here we  have assumed that the central region is surrounded by  an empty  vacuole, however it could be assumed as well that  the central region is filled with a fluid endowed with a different type of symmetry.

Once  the set of equations  describing the dynamics of hyperbolically symmetric fluids have been obtained, we have presented a selection of  exact solutions. These were found under two types of conditions. A first class of solutions emerge from the condition of the vanishing complexity factor defined in \cite{c1} ($Y_{TF}=0$) and the quasi--homologous evolution  (\ref{qsh4}) as defined in \cite{20}. A second class of solutions was found  by looking for models characterized by non--vanishing shear, inhomogeneous energy--density and vanishing four--acceleration (geodesics). This class of solutions are entitled to be considered as  hyperbolically symmetric versions of LTB space--times.

For the first class of models  two exact solutions were found in the  the non--dissipative case. One of them, (\ref{nds1})--(\ref{nds2}), describes a fluid distribution satisfying conditions ${\cal E}=\sigma=0=\Pi=\mu^\prime=0$, which is a reminiscence of the usual FRW space--time. However, unlike the latter it is not geodesic. The second one appears from the geodesic condition,  then the quasi--homologous evolution  becomes homologous, and the solution is described by (\ref{YTF0A1})--(\ref{4b}). This is a geodesic fluid, satisfying also the conditions ${\cal E}=\sigma=0$, and therefore is also a good candidate to be regarded as a  hyperbolical version of the FRW space--time, however unlike the latter, it is anisotropic in the pressure and inhomogeneous in the energy--density. 

Dissipative models  belonging to this first class  of solutions may be found in \cite{hd}.

Among the second class of solutions we have presented first, solutions corresponding  to non--dissipative dust configurations. Comparing with the spherically symmetric case we observe that only one family of solutions   ($k(r)>0$) exists, instead of the three  families existing in this latter case ($k(r)\lesseqqgtr 0$). 

These solutions cannot satisfy the vanishing complexity factor, neither can they evolve in the quasi--homologous regime. Also, the scalar expansion is positive as expected for pure dust submitted to a repulsive gravity.

Next we have analyzed the case of dissipative anisotropic fluids. For doing this we generalized the expression (\ref{BTBh})  by assuming (\ref{ecB}). Different specific models were found from two different conditions. One family of solutions was obtained from a condition imposed on the complexity factor (\ref{ecK}). In this case   the pressure must be anisotropic. A specific solution  was found assuming $P_\bot=0$. The subclass of this type of solution satisfying the vanishing complexity factor is necessarily non--dissipative, and satisfies the stiff equation of state $P_r=-\mu$.

The other family  of solutions was found under the condition $B=1$. For the non--dissipative case a family of solutions was found under the additional condition $P_\bot=0$. In this case too, the vanishing complexity factor condition implies the stiff equation of state $P_r=-\mu$. In the dissipative case,  assuming (\ref{se18}) we may identify two types of contributions in the expression of the temperature. On the one hand   contributions in the stationary dissipative regime (non containing $\tau$)  and on the other hand, contributions from the transient regime (terms proportional to $\tau$).
If we assume further the vanishing complexity factor condition then the fluid becomes non--dissipative and satisfies the  stiff equation of state $P_r=-\mu$. 

The expansion scalar  for all the above models are always positive.

Finally we would like to conclude with some general remarks:
\begin{itemize}

\item All the models exhibited above, although not directly related to   specific scenarios, suggest their potential applications in cosmology and astrophysics. We have in mind the construction of more sophisticated models of the Universe  beyond the  standard FRW  (see for example \cite{cm1,cm2}). More specifically, the repulsive character of gravitation exhibited in hyperbolically symmetric fluids, suggests its possible use in the modeling of very early stages of the Universe.

\item The particular behavior of test particles  within the horizon, assuming the line element (\ref{w3}), suggests a possible model of relativistic jets (see \cite{2nc} for a discussion on this issue).

\item It is worth mentioning that, unlike the spherically symmetric case, there is not a Birkhoff  theorem for hyperbolical symmetry, and therefore neither is the metric (\ref{w3}) unique, nor is the line element (\ref{1}) the most general to describe hyperbolically symmetric fluids. It would be very interesting to delve deeper into this issue, considering other hyperbolically symmetric space--times.
\end{itemize}

\section*{Acknowledgments}

This work was partially supported by Ministerio de Ciencia, Innovacion y Universidades. Grant number:
PGC2018096038BI00. I also wish to thank  Universitat de les  Illes Balears  for financial support and hospitality.

\appendix

\section{Conservation laws}

The two independent components of the conservation laws $T^{\mu \nu}_{;\mu}=0$, read

\begin{equation}\label{mupunto}
  \dot{\mu}+(\mu+P_r)\frac{\dot B}{B}+2(\mu+ P_\bot) \frac{\dot R}{R}+q^\prime \frac{A}{B}+2q\frac{A}{B}\left (
  \frac{A^\prime}{A}+\frac{R^\prime}{R}\right )=0,
\end{equation}

and

\begin{equation}\label{pprima}
  P_r^\prime+(\mu+P_r)\frac{A^\prime}{A}+2(P_r-P_\bot)\frac{R^\prime}{R}+\dot{q}
  \frac{B}{A}+2q\frac{B}{A}\left(\frac{\dot B}{B}+\frac{\dot R}{R}\right)=0.
\end{equation}

This last equation may be rewritten as 
\begin{eqnarray}\label{pprimaU2}
  (\mu+P_r)D_T U&=&-(\mu+P_r)(4\pi P_r R^3-m)\frac{1}{R^2}\nonumber \\&-&E^2 \left[D_R P_r+\frac{2}{R}(P_r-P_\bot)\right]
  \nonumber \\&-&E\left[D_T q+\frac{2}{3}q(2\Theta+\sigma) \right].
\end{eqnarray}

Combining the above equation with the transport equation (\ref{ecTra})  one obtains 
\begin{eqnarray}\label{pprimaU3}
 &&\left(\mu+P_r-\frac{\kappa T}{\tau}\right)D_T U =\nonumber \\ &-&\left(\mu+P_r-\frac{\kappa T}{\tau}\right)\left(4\pi R^3P_r-m\right) \frac{1}{R^2}\nonumber \\
  &- &E^2\left [D_R P_r+\frac{2}{R}(P_r-P_\bot)-\frac{\kappa}{\tau}D_RT\right ]
 \nonumber \\ &+&Eq\left [\frac{1}{\tau}+\frac{1}{2}D_T\ln \left(\frac{\tau}{\kappa T^2}\right)-\frac{5}{6}\Theta-\frac{2}{3}\sigma\right ].
\end{eqnarray}

\end{document}